\documentclass[prl,twocolumn,floatfix,10pt,aps]{revtex4-1}
\pdfoutput=1

\usepackage{amsfonts}
\usepackage{amsmath}
\usepackage{amsthm}
\usepackage{amscd}
\usepackage{eucal}
\usepackage{graphicx}
\usepackage{booktabs}
\usepackage{microtype}
\usepackage{hyperref}
\usepackage{times}

\let\a=\alpha   \let\de=\delta

  \let\p=\pi \let\r=\rho 
\let\om=\omega 
\let\ph=\varphi   
  
  \let\D=\Delta

\let\qd=\quad  

\def\epp{\, .}
\def\epc{\, ,}

\theoremstyle{plain}

\newtheorem*{corollary*}{Corollary}

\newtheorem*{conjecture*}{Conjecture}

\theoremstyle{definition}

\def\2{\frac{1}{2}} \def\4{\frac{1}{4}}

\def\6{\partial}

\def\+{\dagger}

\def\<{\langle} \def\>{\rangle}

\def\i{{\rm i}}

\def\rd{{\rm d}}
\def\re{{\rm e}}

\DeclareMathOperator{\tgh}{th}

\DeclareMathOperator{\tr}{tr}

\DeclareMathOperator{\ad}{ad}

\def\ev{\mathbf{e}}

\def\hv{\mathbf{h}}

\def\Sv{\mathbf{S}}

\begin{document}
\title{On the theory of microwave absorption by the spin-1/2 Heisenberg-Ising magnet}
\author{Michael Brockmann}
\author{Frank G\"ohmann}
\author{Michael Karbach}
\author{Andreas Kl\"umper}
\affiliation{Fachbereich C -- Physik, Bergische Universit\"at Wuppertal, 42097
Wuppertal, Germany}
\author{Alexander Wei{\ss}e}
\affiliation{Max-Planck-Institut f\"ur Mathematik, P.O. Box 7280,
53072 Bonn, Germany}
\begin{abstract}
We analyze the problem of microwave absorption by the Heisenberg-Ising magnet
in terms of shifted moments of the imaginary part of the dynamical susceptibility.
When both, the Zeeman field and the wave vector of the incident microwave, are
parallel to the anisotropy axis, the first four moments determine the shift of
the resonance frequency and the line width in a situation where the frequency is
varied for fixed Zeeman field. For the one-dimensional model we can calculate the
moments exactly. This provides exact data for the resonance shift and the line
width at arbitrary temperatures and magnetic fields. In current ESR experiments
the Zeeman field is varied for fixed frequency. We show how in this situation
the moments give perturbative results for the resonance shift and for the integrated
intensity at small anisotropy as well as an explicit formula connecting the line
width with the anisotropy parameter in the high-temperature limit.
\end{abstract}

\maketitle

The total magnetization in an isotropic system of interacting spins rotates as a
whole about the axis of a homogeneous external field (see e.g.\ \cite{OsAf99,%
*OsAf02}). We consider $L$ spins-$\2$, combining to a total spin $\Sv = S^x \ev_x
+ S^y \ev_y + S^z \ev_z$, $S^\a = \sum_{j=1}^L s_j^\a$, in a magnetic
field of strength $h$ in $z$-direction. The Heisenberg equation of motion for $\Sv$
with a Zeeman term $- h S^z$ is solved by $\Sv (t) = ( \cos(ht) S^x +
\sin(ht) S^y) \ev_x - (\sin(ht) S^x - \cos(ht) S^y) \ev_y + S^z \ev_z$, a rotation
counterclockwise about the $z$-axis. In ESR experiments this can be probed
by circularly polarized microwave radiation propagating along the $z$-direction.
Since its wavelength is large compared to typical distances in spin systems,
we may assume a magnetic field component of the form $\hv (t) = A ( \cos (\om t) \ev_x
- \sin(\om t) \ev_y)$, $A > 0$. It couples to the total spin as $V(t) =
- h^\a (t) S^\a$ and produces a sharp resonance at $\om = h$ (see (\ref{intxxx})
below). If the spin system is perturbed by anisotropic interactions, this resonance
is broadened, shifted or even split in a way that is characteristic of the microscopic
interactions between the spins.

For any spin system with Hamiltonian $H$ linear response theory relates the observed
absorbed intensity to the (imaginary part of the) dynamical susceptibility
\cite{KuTo54}
\begin{equation} \label{defchi}
     \chi_{+-}'' (\om, h) = \frac{1}{2L} \int_{- \infty}^\infty \rd t \:
        \re^{\i \om t} \bigl\< [S^+ (t), S^-] \bigr\>_T \epp
\end{equation}
Here $S^\pm = S^x \pm \i S^y$ and $\< \cdot \>_T$ stands for the canonical average
at temperature $T$ calculated by means of the statistical operator $\r =
\re^{- (H - h S^z)/T}/\tr \re^{- (H - h S^z)/T}$. Through this average the dynamical
susceptibility depends on $h$ and $T$. The absorbed intensity per spin, normalized by the
intensity $A^2$ of the incident wave and averaged over a half-period $\p/\om$ of the
microwave field, is
\begin{equation} \label{int}
     I (\om, h) = \frac\om2 \chi_{+-}'' (\om, h) \epp
\end{equation}

In current ESR experiments in solids $I (\om, h)$ is measured as a function of $h$
for fixed $\om$. Of particular interest are experiments on quasi
one-dimensional compounds (reviewed e.g.\ in \cite{Ajiro03,KBL10}) which provide
prototypical realizations of interacting many-body systems with strong quantum
fluctuations. Still, the data are not always easy to interpret, because of a lack
of reliable theoretical predictions.

Most of the existing theories are based on a priori assumptions about the line shape
and typically apply for limited ranges of temperature and magnetic field.
Field theoretical approaches \cite{OsAf99,*OsAf02} are restricted to small
temperatures and small (but not too small) magnetic fields. The more traditional
approaches \cite{KuTo54,NaTa72} rely on the high temperature approximation. Purely
numerical approaches \cite{MYO99,*OgMi03,*ECM10} are unbiased, but the extrapolation
of the data to the thermodynamic limit of large chains may be difficult.

A remarkable result for the resonance shift in one-dimensional antiferromagnetic
chains, being valid at arbitrary temperatures, was obtained in \cite{MSO05}.
It utilizes the exact nearest-neighbor correlation functions of the isotropic
spin-$\2$ Heisenberg chain. Here we present an alternative framework for the
derivation of the resonance shift which, in the limit of small anisotropy,
reproduces \cite{MSO05}. In our approach the anisotropy is treated
non-perturbatively, and it allows us to derive an exact formula for the line width
`in frequency direction' at fixed magnetic field, as well as a new explicit
expression for the ESR-line width in the high temperature regime.

We consider an important example of anisotropic interactions described by the
Heisenberg-Ising Hamiltonian
\begin{equation} \label{ham}
     H = J \sum_{\<i j\>} \bigl( s_i^x s_j^x + s_i^y s_j^y + (1 + \de)
                               s_i^z s_j^z \bigr) \epp
\end{equation}
Here the sum is over nearest neighbors, and $\de$ is the anisotropy parameter.
If $\de = 0$, then (\ref{defchi}), (\ref{int}) imply that the
normalized absorbed intensity is 
\begin{equation} \label{intxxx}
     I (\om, h) = \p \de(\om - h) h m(T, h) \epp
\end{equation}
It is proportional to the magnetic energy $h m(T,h)$ per lattice site. This
case includes the familiar paramagnetic resonance (Zeeman effect) for $J = 0$, for
which the magnetization is $m(T,h) = \2 \tgh \bigl( \frac{h}{2T} \bigr)$.

For non-zero $\de$ the function $\chi_{+-}''$ is unknown and hard to calculate.
Still, some more elementary spectral characteristics, such as the position of
the resonance or the line width, may be expressed in terms of certain static
correlation functions that determine the moments of the normalized intensity function.

\begin{figure}
\begin{center}
\resizebox{0.48\textwidth}{!}{\input{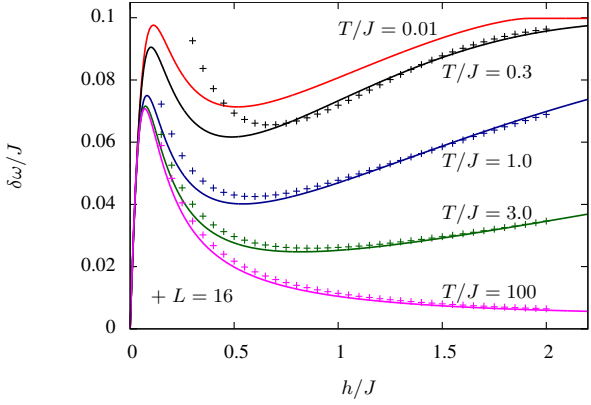}}
\caption{\label{fig:shifthcrit} Resonance shift $\de \om/J$ for the 1D model
in the critical regime at $\de = - 0.1$ as function of the magnetic field. Crosses
from fully numerical calculation for a finite chain Hamiltonian of 16 sites.}
\end{center}
\end{figure}
Let us assume for a while that our chain is large but finite. Then the
spectrum is bounded and the integrals
\begin{equation}
     I_n = \int_{- \infty}^\infty \rd \om \: \om^n I (\om, h)
\end{equation}
exist for all non-negative integers $n$. Since $I (\om, h)$ is non-negative
everywhere and since $I_0 > 0$, we may interpret $I (\om, h)/I_0$ as a
probability distribution and the $I_n$ as its moments. As we shall see, it
is convenient to express the $I_n$ in terms of another closely related sequence
of integrals
\begin{equation} \label{defm}
	m_n (T, h) = J^{-n} \int_{- \infty}^\infty
	                    \frac{\rd \om}{2 \p} (\om - h)^n \chi_{+-}'' (\om, h)
\end{equation}
which, by slight abuse of language, will be called (shifted) moments. Again
they exist for every finite chain.

By definition the shift of the resonance for fixed $h$ is
\begin{equation} \label{rshiftmom}
     \de \om = \frac{I_1}{I_0} - h = J \frac{J m_2 + h m_1}{J m_1 + h m_0} \epp
\end{equation}
A measure for the line width is the mean square deviation
\begin{equation} \label{widthmom}
     \D \om^2 = \frac{I_2}{I_0} - \frac{I_1^2}{I_0^2} =
        J^2 \frac{J m_3 + h m_2}{J m_1 + h m_0} - \de \om^2 \epp
\end{equation}
Hence, in order to calculate the resonance shift and the line width, we need
to know the first four shifted moments $m_0$, $m_1$, $m_2$, $m_3$ of the dynamic
susceptibility $\chi_{+-}''$.

\begin{figure}
\begin{center}
\resizebox{0.48\textwidth}{!}{\input{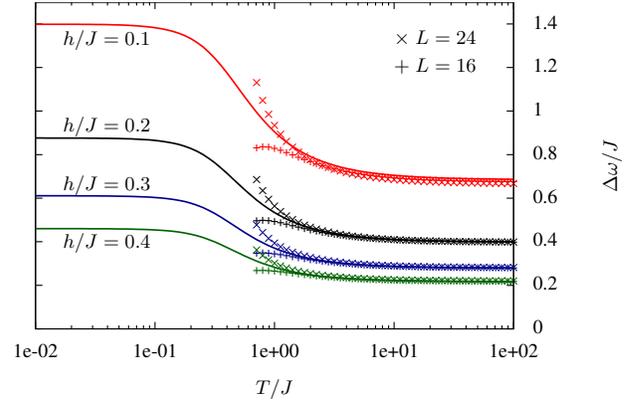}}
\caption{\label{fig:widthcrit} Line width $\D \om/J$ for the 1D model
in the critical regime at $\de = - 0.1$ as function of temperature.
Crosses from fully numerical calculation for finite chain Hamiltonians of 16 and
24 sites.}
\end{center}
\end{figure}
In the following we shall employ the notation $\ad_X \cdot = [X, \cdot]$ for
the adjoint action of an operator $X$. Then $S^+ (t) = \re^{- \i ht}
\re^{\i t \ad_{H}} S^+$, since $[H,S^z] = 0$ and $[S^z, S^+] = S^+$, and
it follows with (\ref{defchi}) and (\ref{defm}) that
\begin{equation} \label{allmoments}
     m_n = \frac1{2L} \bigl\< [S^+, \ad_{H/J}^n S^-] \bigr\>_T \epp
\end{equation}

The latter formula shows that the moments $m_n$ are static correlation
functions whose complexity grows with growing $n$. The first few of them
can be easily calculated. We shall show the results for the one-dimensional
model, for which
\begin{equation} \label{m0}
     m_0 = \frac1{2L} \bigl\< [S^+, S^-] \bigr\>_T
         = \frac1{L} \bigl\< S^z \bigr\>_T \epc
\end{equation}
which is the magnetization per lattice site. The subsequent moments are less
intuitive,
\begin{subequations}
\label{m13}
\begin{align} \label{m1}
     m_1 & = \de \< s_1^+ s_2^- - 2 s_1^z s_2^z \>_T \epc \\
     m_2 & = \2 \de^2 \< s_1^z + 4 s_1^z s_2^z s_3^z - 4 s_1^z s_2^+ s_3^- \>_T
               \epc \displaybreak[0] \\
     m_3 & = \4 \de^2 \bigl\< 2 s_1^+ s_2^+ s_3^- s_4^-
                          + 4 s_1^+ s_2^- s_3^+ s_4^-
                          - 2 s_1^+ s_2^- s_3^- s_4^+ \notag \\ & \mspace{27.mu}
                          - 8 s_1^z s_2^z s_3^+ s_4^-
                          - 4 s_1^z s_2^+ s_3^z s_4^-
                          + 8 s_1^z s_2^+ s_3^- s_4^z
			  - 4 s_1^+ s_2^- \notag \\ & \mspace{27.mu}
			  - s_1^+ s_3^- + 8 s_1^z s_2^z s_3^z s_4^z
			  + 2 s_1^z s_3^z - 4 s_1^z s_2^z \notag \\ & \mspace{27.mu}
			  + \de ( 8 s_1^z s_2^+ s_3^- s_4^z + 2 s_1^+ s_2^-
			  - 8 s_1^z s_2^z ) \bigr\>_T \epp
\end{align}
\end{subequations}
They are certain combinations of static short-range correlation functions.
This implies, in particular, that they all exist in the thermodynamic limit
$L \rightarrow \infty$. It follows that the line shape for fixed $h$ cannot
be strictly Lorentzian, as is often assumed in the literature.

All static correlation functions of the one-dimensional Heisenberg-Ising
model are polynomials in the derivatives of three functions $\om$, $\om'$
and $\ph$ \cite{JMS08} which, as is common in integrable models, can be
expressed in terms of the solutions of certain well behaved linear and
non-linear integral equations \cite{BoGo09}. This is the reason why in
this case the moments $m_0, m_1, m_2, m_3$ can be calculated exactly
by means of the techniques developed in \cite{BGKS07,*BDGKSW08,*TGK10a}. 

We parameterize the anisotropy as $\de = (q - 1)^2/2q$. Then, with
the shorthand notations $\ph_{(n)} = \6_x^n \ph (x)|_{x=0}$, $f_{(m,n)} =
\6_x^m \6_y^n f(x, y)|_{x=y=0}$, for $f = \om, \om'$, we obtain
\begin{widetext}
\begin{align} \label{ms}
     m_0 & = - \2 \ph_{(0)} \epc \qd 
     m_1 = \frac{(q - 1)^2 (q^2 + 4q + 1) \om_{(0,1)}'}{16q^2}
             - \frac{(q^3 - 1) \om_{(0,0)}}{4 q (q + 1)} \epc \notag \\
     m_2 & = \frac{(q - 1)^2}{256 q^4} \bigl[
             4 q (q + 1)(q^3 - 1)(\om_{(0,2)} \ph_{(0)}
	                - 2 \om_{(1,1)} \ph_{(0)} - \om_{(0,0)} \ph_{(2)}) \notag \\
         & \mspace{126.mu} + (q^2 - 1)^2 (q^2 + 4q + 1)
	   (\om_{(1,2)}' \ph_{(0)} + \om_{(0,1)}' \ph_{(2)})
	   - 16 q^2 (q - 1)^2 \ph_{(0)} \bigr] \epc \displaybreak[0] \notag \\
     m_3 & = \frac{(q - 1)^4}{98304 q^8 (q^4 - 1)(q^6 - 1)} \notag \\ & \bigl[
             16 q^2 (q^2 - 1)^3 (q^4 - 1)(q^6 - 1)(q^2 + 4q + 1)
	     (\om_{(0,2)} \om_{(0,1)}' + \om_{(0,0)} \om_{(1,2)}') \notag \\ &
	   + 64 q^2 (q^2 - 1)^4 (2 q^{10} - q^9 + 4 q^8 - 4 q^7 - 12 q^6
	             - 14 q^5 - 12 q^4 - 4 q^3 + 4 q^2 - q + 2)
		     \om_{(0,0)} \om_{(1,1)} \notag \displaybreak[0] \\ &
           - 16 q^2 (q^2 - 1)^2 (q^4 - 1)(q^6 - 1)(3 q^4 + 14q^2 + 3) \om_{(0,3)}'
	     \notag \\ &
           + 8 q^2 (q^2 - 1) (q^4 - 1)^2 (q^6 - 1) (q + 1)^2
	           (8 \om_{(1,2)}' - \om_{(2,3)}') \notag \\ &
           + 192 q^4 (q^2 - 1)^2 (q^4 - 1) (q^6 + 18 q^4 + 8 q^3 + 18 q^2 + 1)
	              \om_{(0,2)} \notag \displaybreak[0] \\ &
           + 64 q^2 (q^2 - 1)^2 (q^4 - 1) (q + 1)^2 (2 q^8 - 5 q^7 + 26 q^6 - 49 q^5
	             + 28 q^4 - 49 q^3 + 26 q^2 - 5 q + 2) \om_{(1,1)} \notag \\ &
           - 16 q^2 (q^2 - 1)^2 (q^4 - 1) (q + 1)^2 (q^8 - q^7 + q^6 + q^5 + 2 q^4
	             + q^3 + q^2 -  q + 1) (2 \om_{(1,3)} - 3 \om_{(2,2)}) \notag
		     \displaybreak[0] \\ &
           + 64 q^2 (q^4 - 1) (q^6 - 1) (3 q^8 + 2 q^6 + 24 q^5 - 130 q^4 + 24 q^3
	             + 2 q^2 + 3) \om_{(0,1)}' \notag \displaybreak[0] \\ &
           + (q^4 - 1) (q^6 - 1) (q + 1)^2 (q^{10} - 2 q^9 + 25 q^8 + 16 q^7 + 118 q^6
	             + 164 q^5 \notag \\ & \mspace{126.mu} 
		     + 118 q^4 + 16 q^3 + 25 q^2 - 2 q + 1)
	     (\om_{(0,3)}' \om_{(1,2)}' + \om_{(0,1)}' \om_{(2,3)}') \notag \\ &
           - 1536 q^5 (q^4 - 1) (4 q^8 - 9 q^7 - 2 q^6 - 6 q^5 + 8 q^4 - 6 q^3
	             - 2 q^2 - 9 q + 4) \om_{(0,0)}
		     \displaybreak[0] \notag \\ &
           + 4 q^2 (q^6 - 1) (q + 1)^2 (q^2 + 1) (5 q^8 - 2 q^7 + 32 q^6 + 50 q^5
	             + 70 q^4 + 50 q^3 + 32 q^2 - 2 q + 5) \notag \\ & \mspace{54.mu}
	     (2 \om_{(1,3)} \om_{(0,1)}' - 3 \om_{(2, 2)} \om_{(0,1)}'
	      + \om_{(0,2)} \om_{(0,3)}' - 2 \om_{(1,1)} \om_{(0,3)}'
	      - 3 \om_{(0,2)} \om_{(1,2)}' - \om_{(0,0)} \om_{(2,3)}')
	      \notag \displaybreak[0] \\ &
           - 16 q^2 (q + 1)^2 (q^{16} - q^{15} + 8 q^{14} + 9 q^{13} + 47 q^{12}
	             + 45 q^{11} + 96 q^{10} + 91 q^9 + 128 q^8 + 91 q^7 + 96 q^6
		     \notag \\ & \mspace{72.mu}
		     + 45 q^5 + 47 q^4 + 9 q^3 + 8 q^2 - q + 1)
	     (3 \om_{(0,2)}^2 - 6 \om_{(1, 1)} \om_{(0,2)}
	      + 2 \om_{(0,0)} \om_{(1,3)} - 3 \om_{(0,0)} \om_{(2,2)}) \bigr] \epp
\end{align}
\end{widetext}
These functions represent the moments in the thermodynamic limit. Since
they can be calculated to arbitrary precision, we obtain numerically
accurate results for the resonance shift and for the line width as functions
of temperature or magnetic field over the whole range of the phase diagram. In
particular, our approach is not restricted to small anisotropies. Examples for
$\de = - 0.1$ are shown in figures \ref{fig:shifthcrit} and \ref{fig:widthcrit}.
We find a broadening of the line width as defined by (\ref{widthmom}) for small
temperatures in the critical ($\de < 0$) as well as in the massive
($\de > 0$) regime (latter case not shown here).

At first sight this seems to contradict experimental results \cite{KBL10} which
claim a narrowing. Still, one has to take into account that usually in the 
analysis of experimental data rather different definitions of the line width, 
as e.g. the distance between the turning points right and left to the maximum of
the intensity, are used. In particular, if the intensity distribution has long
shallow tails the definition (\ref{widthmom}) will give considerably larger values
than the distance between the turning points. We have performed a detailed numerical
study of the dynamical susceptibility for chains of 16, 20 and 24 lattice sites
(to be published elsewhere) and we see indeed such long tails at low temperature
(compare also \cite{ECM10}). In experiments they may be misinterpreted as
background stemming from couplings of the spin chain to other degrees of freedom,
but in fact they are due to the spin-spin interactions and are part of the true
absorption line. For the determination of the resonance shift tails are expected
to have less influence. As long as they are not too much asymmetric the shift
calculated by means of (\ref{rshiftmom}) should agree with the shift of the
maximum of the absorbed intensity.

In current ESR experiments the microwave frequency $\om$ is kept fixed and the
Zeeman field $h$ is modulated. This means that, as opposed to most of the
theoretical treatments, including our considerations above, the absorbed intensity
$I (\om, h) = \om \chi_{+-}'' (\om, h)/2$ is determined as a function of $h$
for fixed $\om$, and the resonance shift and line width are measured in
`$h$-direction'. Away from the isotropic point ($\de = 0$), where
$\chi_{+-}'' (\om, h)$ is symmetric and the absorption line is extremely narrow,
this may clearly lead to rather different values. To be closer to present-day ESR
experiments one should calculate resonance shift and line width in terms of the
moments of the dynamical susceptibility in `$h$-direction'.

We define
\begin{equation}
	M_n (T, \om) = J^{-n} \int_{- \infty}^\infty
	               \frac{\rd h}{2 \p} (h - \om)^n \chi_{+-}'' (\om, h) \epp
\end{equation}
For these functions we obtain the representation
\begin{equation} \label{Mom}
     M_n (T, \om) = (- 1)^n
        \sum_{k=0}^\infty \frac{(- J)^k}{k!} m_{k + n}^{(k)} (T, \om) \epc
\end{equation}
where the superscript $(k)$ denotes the $k$th derivative with respect to the second
argument. We see that the $M_n$ are determined by infinitely many of the $m_n$ and
their derivatives, i.e., they depend on static correlation functions for arbitrarily
large distances. For this reason they cannot be calculated by our exact method
above. Yet, in certain cases only finitely many terms of the series are needed for
a good approximation.

We first of all express the resonance shift $\de h = \<h\> - \om$ and the mean
square deviation from the center of the absorption peak $\D h^2 = \<h^2\> - \<h\>^2$
in terms of the $M_n$,
\begin{equation} \label{shiftwidthh}
     \frac{\de h}{J} = \frac{M_1}{M_0} \epc \qd
     \frac{\D h^2}{J^2} = \frac{M_2}{M_0} - \frac{M_1^2}{M_0^2} \epp
\end{equation}
We have identified two cases, where these formulae simplify and finitely many
of the $m_n$ are enough to determine $\de h$ and $\D h$ approximately.

The equation for the resonance shift simplifies for small anisotropy $|\de| \ll 1$.
Since $M_0 = m_0 + {\cal O} (\de)$, $M_1 = m_1 + {\cal O} (\de^2)$ and, generically,
$m_1$ itself is of order $\de$ (see (\ref{m13}), (\ref{Mom})) we obtain to linear
order in~$\de$
\begin{equation} \label{hshiftapp}
     \frac{\de h}{J} = - \frac{m_1}{m_0} \epp
\end{equation}
In \cite{NaTa72,MSO05} the same equation was obtained by a more intuitive reasoning.
It leads to results which compare rather well to experiments \cite{MSO05}. However,
some care is necessary with the interpretation of (\ref{hshiftapp}). $m_1/\de$ vanishes
at $\de = h = 0$. It follows that $m_1 = \de (a h + b \de + \dots)$ with some
coefficients $a, b$, whence $h$ must be large compared to $\de$ for (\ref{hshiftapp})
to be applicable. Note that all higher moments $m_n$ are of order $\de^2$. Hence,
there is no simplification for small anisotropy, like in (\ref{hshiftapp}), for the
line width. But the integrated intensity $M_0$ has again a finite approximation to first
order in $\de$, $M_0 = m_0 - J m_1'$.

The representation (\ref{Mom}) is a series in ascending powers of $J/T$ (with
still temperature dependent coefficients). This can be used to evaluate 
(\ref{shiftwidthh}) asymptotically for high temperatures. It turns out that
that the leading terms in the $J/T$ expansion of $m_1$ and $J m_2'$ cancel each
other: $\de h \sim \frac{h}{2T} \de \rightarrow 0$ in the high-temperature
limit $T \gg J$, and
\begin{equation} \label{highwidth}
     \frac{\D h}{J} = \frac{|\de|}{\sqrt{2}}
\end{equation}
for arbitrary microwave frequency $\om$. This formula provides a simple means
to directly measure the anisotropy parameter $\de$.

It may be instructive to illustrate our formula with one of the few
explicit results, namely with the formula for the intensity in the free
fermion case $\de = -1$ at $T \rightarrow \infty$ \cite{BrJa76}. In this case
$I (\om, h) \sim (\om^2/J^2) \exp \bigl(- (\om - h)^2/J^2\bigr)$ and,
in agreement with (\ref{highwidth}), we obtain the line width $\frac{\D h}{J} =
\frac{1}{\sqrt{2}}$, whereas the width in omega direction depends on~$h$.

Our work is the first exact result for the resonance shift and the line width in
microwave absorption experiments on the Heisenberg-Ising chain. The reduction to
moments is not restricted to the integrable case and may be interesting for the two-
and three-dimensional models as well. Our approach is unbiased. It makes no
a priori assumptions about the shape of the spectral line. As opposed to all
other approaches it is valid for all temperatures and magnetic fields and in
addition for arbitrary values of $\de$. For small $\de$ close to the isotropic
point we recover the result of \cite{MSO05} for the line shift. The resonance shift
$\de \om/J$ or $\de h/J$ and the line width $\D \om/J$ or $\D h/J$ defined in
terms of moments show a simple scaling behavior. They depend on the exchange
interaction only through the ratios $T/J$ and $h/J$. In this sense the curves in
figures \ref{fig:shifthcrit} and \ref{fig:widthcrit} are universal.

The intensity $I (\om, h)$ is a function of $\om$ and $h$. With our definitions of
$\de \om$ and $\D \om$ we determine the resonance shift and the line width in
$\om$-direction as functions of $h$, while in standard ESR experiments the resonance
shift $\de h$ and line width $\D h$ in $h$-direction are measured as functions
of $\om$, which should be clearly distinguished. For the resonance shift it follows
from (\ref{rshiftmom}), (\ref{hshiftapp}) that $\de h (T, \om) =
- \de \om (T, h)|_{h = \om}$ to linear order in $\de$. For the line width there is
no such simple relation between $\D \om$ and $\D h$. However, for $\D h$ we obtained
the simple high-temperature formula (\ref{highwidth}) which we suggest to be useful
to measure the anisotropy directly. We are further convinced that it may be worth
trying to measure $\D \om$, which is now known exactly, directly in multi-frequency ESR
experiments.

\begin{acknowledgments}
The authors would like to thank F. Anders, H. Bomsdorf, H. Boos, B. Lenz, K. Sakai,
J. Stolze, A. Zvyagin, and, in particular, Y. Maeda for stimulating discussions.
\end{acknowledgments}


%

\end{document}